\documentclass[twocolumn,aps,prl,10pt,amsmath,amssymb,nofootinbib,showpacs,superscriptaddress,floatfix]{revtex4-2}

\newcommand{\rs}{\rm\scriptscriptstyle}

\DeclareFontFamily{U}{rcjhbltx}{}
\DeclareFontShape{U}{rcjhbltx}{m}{n}{<->rcjhbltx}{}
\DeclareSymbolFont{hebrewletters}{U}{rcjhbltx}{m}{n}

\usepackage{graphicx}
\usepackage{color}
\usepackage[usenames,dvipsnames]{xcolor}
\usepackage[colorlinks=true,linkcolor=Red,citecolor=Green,linktoc=page]{hyperref}
\usepackage{multirow}
\usepackage{float}
\usepackage{flushend}
\usepackage{balance}
\usepackage[varg]{txfonts}
\usepackage{ulem}
\usepackage{fancyhdr}

\DeclareMathSymbol{\lamed}{\mathord}{hebrewletters}{108}

\begin{document}
\title{Type-III Superconductivity}

\author{M.\,C.\,Diamantini}

\affiliation{NiPS Laboratory, INFN and Dipartimento di Fisica e Geologia, University of Perugia, via A. Pascoli, I-06100 Perugia, Italy}

\author{C.\,A.\,Trugenberger}

\affiliation{SwissScientific Technologies SA, rue du Rhone 59, CH-1204 Geneva, Switzerland}

\author{Sheng-Zong Chen}

\affiliation{Department of Physics, National Taiwan University, Taipei 106, Taiwan}

\author{Yu-Jung Lu}

\affiliation{Research Center for Applied Sciences, Academia Sinica, Taipei 115, Taiwan and Department of Physics, National Taiwan University, Taipei 106, Taiwan}

\author{Chi-Te Liang}

\affiliation{Department of Physics, National Taiwan University, Taipei 106, Taiwan and Center for Quantum Science and Engineering, 
National Taiwan University, Taipei 106, Taiwan}

\author{V. M. Vinokur}
\affiliation{Terra Quantum AG, St. Gallerstrasse 16A, 9400 Rorschach, Switzerland}

\
\begin{abstract}
Superconductivity remains one of most fascinating quantum phenomena existing on a macroscopic scale. Its rich phenomenology is usually described by the Ginzburg-Landau (GL) theory in terms of the order parameter, representing the macroscopic wave function of the superconducting condensate. The GL theory addresses one of the prime superconducting properties, screening of the electromagnetic field because it becomes massive within a superconductor, the famous Anderson-Higgs mechanism. Here we describe another widely-spread type of superconductivity where the Anderson-Higgs mechanism does not work and must be replaced by the Deser-Jackiw-Templeton topological mass generation and, correspondingly, the GL effective field theory must be replaced by an effective topological gauge theory. These superconductors are inherently inhomogeneous granular superconductors, where electronic granularity is either fundamental or emerging. We show that the corresponding superconducting transition is a three-dimensional (3D) generalization of the 2D Berezinskii-Kosterlitz-Thouless (BKT) vortex binding-unbinding transition. The binding-unbinding of the line-like vortices in 3D results in the Vogel-Fulcher-Tamman (VFT) scaling of the resistance near the superconducting transition. We report experimental data fully confirming the VFT behavior of the resistance.	
\end{abstract}

\maketitle

\section{Introduction}
	The macroscopic physics of traditional superconductors (SC) is governed by the Ginzburg-Landau (GL) model, see e.g.,\,\cite{tinkham}, describing superconductors in terms of a local order parameter. The ground state of a macroscopic superconductor of a size much exceeding the London penetration depth, $\lambda_{\rs L}$, has an order parameter that is constant in the system's bulk, outside a boundary shell of the width $\lambda_{\rs L}$. There are, however, superconductors, for example thin films of a thickness $d$ comparable with the coherence length $\xi$, which are characterized by a completely different ground state exhibiting a paradigmatic granularity\,\cite{granular, bat_vin_2007}. In these systems superconductivity sets in when global phase coherence is established due to tunneling Cooper pairs percolation between droplets of locally formed condensate. The granularity of these systems is associated with a superconductor-to-superinsulator quantum phase transition\,\cite{dst, vinokurnature, dtv1} which may occur via an intermediate Bose metal\,\cite{dst, das1, bm} phase, see\,\cite{dtv-SIT,book1,book}, and has been detected experimentally\,\cite{sacepe1, sacepe2}. 
	
	Two important features characterize these planar ``self-granular" superconductors. First, the gauge screening length, the Pearl length $\lambda_{\rs P}=\lambda_{\rs L}^2/d$ due to the familiar Anderson-Higgs mechanism, see e.g.\,\cite{tinkham}, would become larger than the experimental system size for small $d$. Second, near the quantum transition, the electric fields induced by charges residing in the system remain in-plane over the whole sample because of the large dielectric constant\,\cite{dtv-SIT,book1,book}. Usually, such electric fields are not very relevant in standard superconductors. In our planar superconductors, however, these 2D electric fields are much stronger than the usual 3D ones, since they decay as $1/r$ with the increasing distance $r$ from the charge and cannot be neglected. When coupled to electromagnetism, the time-dependent Ginzburg-Landau model becomes, therefore (non-relativistic) scalar quantum electrodynamics (QED), which is ill defined in 2D because of its infrared divergences tied to the perturbative coupling scaling as ${\rm ln}(L/\xi)$, where $L$ is the sample size\,\cite{polyakov, templeton}. If one tries to derive the free energy for a putative order parameter from the elastic interaction with a substrate one obtains a non-local functional describing a self-organized array of superconducting islands\cite{vinokur2011}, confirming the the original GL model breaks down. Hence, for planar systems with long-range interactions, the local Ginzburg-Landau model does not provide an adequate description of the global superconductivity\,\cite{infrared} and can only address local superconductivity within droplets of the typical size of order $\xi$. 
	
	Notably, emergent granularity is not confined to thin films. Recently, the same physics has been detected in bulk samples\,\cite{3Dgranular}. Even more importantly, the ground-state of the high-$T_c$ cuprates is inhomogeneous, especially in the underdoped regime and the same percolation model is thought to be responsible for global superconductivity \cite{muller, barisic}. In this type of percolating superconductivity, electron pairs survive above $T_c$, which is the case both in 2D\,\cite{pairs2D} and 3D\,\cite{pairs3D}, and there is a quantum transition to an insulating state, both in 2D, see\,\cite{dtv-SIT,book1,book}, and 3D\,\cite{SIT3D}. In 2D, the fragmentation into separate condensate droplets is due to strong infrared divergences near the quantum insulating transition\,\cite{infrared}. In general, however, this type of behaviour seems characteristic of superconductors in which the pairing mechanism is not the BCS one\,\cite{tinkham} but arises from a stronger attractive interaction leading to the pronounced bosonic pairs of a size much smaller than their typical separation distance and forming a Bose-Einstein condensate (BEC). The BECs with long-range dipole interactions of particles carrying magnetic moments are known to fragment into liquid droplets due to strong quantum fluctuations\,\cite{bec}. 
	
	Here we formulate the effective long-distance gauge theory of inhomogeneous superconductors in 3D  and show that superconductivity sets in via the Vogel-Fulcher-Tammann (VFT) transition, the three dimensional counterpart of the 2D Berezinskii-Kosterlitz-Thouless (BKT) topological phase transition, and we report an experiment confirming the VFT scaling of the resistance at the transition. 
	The crucial point is that, contrary to type II superconductors, vortices in inhomogeneous superconductors are not Abrikosov vortices but, rather, mobile vortices with no dissipative core which arise due to non-trivial phase circulations on adjacent droplets. Charges on the droplets and vortices between them have unavoidable topological mutual statistics interactions and a local descriptions of these requires the introduction of gauge fields \cite{wilczek}. Therefore, the effective field theory for these inhomogeneous superconductors must be a gauge theory. We refer to this novel, topologically driven superconducting state (in any dimension) as to type-III superconductivity. 
	This choice is dictated by the standard classification of superconductors with respect to penetration of the applied magnetic field. Type\,I superconductors expel magnetic field $H$ and are referred to be in a Meissner state at $H<H_{\mathrm c}$, while  at the critical field $H_{\mathrm c}$ superconductivity destroys. In type II superconductivity $H_{\mathrm c}$ ``splits" into the lower, $H_{\mathrm c1}$, and upper, $H_{\mathrm c2}$, critical fields. At $H<H_{\mathrm c1}$ type\,II superconductors are in the Meissner state; at $H_{\mathrm c1}<H<H_{\mathrm c2}$ (mixed or vortex state) magnetic field penetrates type\,II superconductors in a form of Abrikosov vortices having a normal core; and at $H=H_{\mathrm c2}$ vortex normal cores overlap and superconductivity gets destroyed. In type\,III superconductors $H_{\rm c1}=0$ and vortices, which, as mentioned above, in this case do not possess normal core, can penetrate at any magnetic field (corresponding to a flux at least equal to a quantum flux), and there is no true Meissner state. Superconductivity is not destroyed at low temperatures because the quantum diffusion of vortices is suppressed by the large corresponding term in the action. This behavior has been experimentally detected\,\cite{para} in Josephson junction arrays, the paradigmatic example of type III superconductors in 2D. The response can be both diamagnetic and paramagnetic, however since it is preferentially paramagnetic this state is sometimes called paramagnetic Meissner state. Finally, in type\,III superconductors, vortices can proliferate even without a magnetic field when the temperature is high enough. 

	%
	


	
	\section{BKT and VFT transitions}
	As we have established in the Introduction, the local GL model fails to provide a consistent description of granular or droplet-composed superconductors\,\cite{barisic} and has thus to be replaced by a generalized gauge theory introduced in\,\cite{topsc, higgsless} and recently discussed in detail in 2D in\,\cite{infrared}. 
	One of the fundamental implications of the gauge theory of granular superconductors is that its vortices, contrary to Abrikosov vortices, have no dissipative core, since they arise from non-trivial circulations of the local phases of the condensate on adjacent droplets. Hence,  superconductivity in these materials that do not possess a global order parameter is referred to as ``Higssless superconductivity"\,\cite{higgsless}. Furthermore, since the ground state of Higgsless superconductors  may carry topological order  they are also called ``superconductors with topological order"\,\cite{topsc}.
	
	Since superconductivity is realized by global phase coherence establishing over all pre-existing condensate droplets, its destruction is caused by a proliferation of vortices and not by breaking of Cooper pairs, as in traditional superconductors. In 2D, this is the famed BKT transition, see\,\cite{minnhagen} for a review, resulting in the BKT resistance scaling 
	\begin{equation}
		R (T) \propto {\rm e}^{-\sqrt{b\over |T-T_{BKT}|}} \ ,
		\label{bkt}
	\end{equation}
	where $b$ is a constant having the dimensionality of temperature. In 3D, the phase transition is again caused by vortex liberation, but the vortices are now 1D extended objects, magnetic strings. The superconducting phase transition is thus caused by 1D strings becoming tensionless. This transition has been studied in\,\cite{dtv2}. The corresponding behaviour of the resistance is modified to the VFT scaling
	\begin{equation}
		R (T) \propto {\rm e}^{-{b^{\prime}\over |T-T_{VFT}|}} \ .
		\label{vft}
	\end{equation}
	This same dual scaling, for vanishing conductivity and due to electric strings becoming tensionless, see\,\cite{book} for a review, has already been detected at the superinsulating side of the quantum transition\,\cite{shahar} and has also been obtained in the XY model with quenched disorder, which apparently is equivalent to one more effective space dimension\,\cite{vasin,disBKT}. 
	
	\section{Gauge theory of type-III superconductors}
	We model an inhomogeneous superconductor by a cubic lattice, with the sites representing the droplets and the links encoding possible tunneling junctions between them. The fundamental degrees of freedom of the model are of two types. First, there are
	integer (in units of 2e) charges $Q_0$ located at the sites and currents $Q_i$ on the links of lattice. Together they constitute a four-current $Q_{\mu}$, with Greek letters standing for the space-time indices. In the absence of background charges, this current is conserved, $d_{\mu} Q_{\mu} = 0$ with $d_{\mu}$ denoting the forward lattice derivative in the direction $\mu$; summation over equal Greek indices is implied. Conservation requires that only closed loops $Q_{\mu}$ are allowed on the lattice, representing charge-hole fluctuations. The second type of excitations are integer (in units $2\pi/2e$, we use natural units $c=1$, $\hbar =1$, $\varepsilon_0=1$) coreless Josephson vortices that arise from the nontrivial circulations of the local condensate phases on the droplets. Since these circulations are 1D extended objects they are represented by integer lattice plaquette variables $M_{\mu \nu}$. Since the vortices that we consider are closed loops, these are also conserved, $d_{\mu} M_{\mu \nu} = d_{\nu} M_{\mu \nu} =0$ and describe, thus, closed surfaces representing the nucleation and subsequent annihilation of a vortex loop. Open vortices with magnetic monopoles at their ends are also possible\,\cite{moncon} but are not relevant for what follows. 
	
	The infrared (IR) dominant interaction between charges and vortices is topological, it encodes their mutual statistical interaction i.e., the Aharonov-Bohm\,\cite{bohm} and Aharonov-Casher\,\cite{casher} (ABC) phases accumulating when one type of excitation encircles the other. In the Euclidean partition function, which we will consider in the rest of this paper, the ABC topological interactions are accounted for by an imaginary term representing the Gaussian linking of the closed loops and surfaces in four Euclidean dimensions\,\cite{kaufmann}. As pointed out by Wilczek\,\cite{wilczek}, this interaction can also be represented in local form by introducing two fictitious gauge fields interacting with the two types of excitations and with a topological coupling to each other. In the 2D this is the well known Chern-Simons term\,\cite{jackiw2}; in the 3D case, this is the three-index BF term\,\cite{book}, $\epsilon_{\mu \nu \alpha \beta} \partial_{\nu} $, coupling an effective gauge field $a_{\mu}$ for the charges with the two-index effective gauge field $b_{\alpha \beta}$ for the 1D extended vortices. Here $\epsilon^{\mu \nu \alpha \beta}$ is the usual totally antisymmetric tensor. Then the Euclidean action acquires the form
	\begin{equation}
		S = \sum_x i {\ell ^4\over 4\pi }
		a_{\mu }k_{\mu \alpha \beta }b_{\alpha \beta } 
		+i\ell a_{\mu }Q_{\mu }
		+i\ell ^2{1  \over 2} b_{\mu \nu}M_{\mu \nu}\,, 
		\label{ac1}
	\end{equation}
	where $x$ denotes the lattice sites, $\ell$ is the link length and $k_{\mu \alpha \beta} $ is the lattice BF term, described in detail in  Methods. 
	
	The two gauge fields are invariant under the gauge transformations
	\begin{eqnarray}
		a_{\mu}\to a_{\mu} + d_{\mu} \xi \ ,
		\nonumber \\
		b_{\mu \nu}\to b_{\mu \nu} +d_{\mu} \lambda_{\nu} -d_{\nu} \lambda_{\mu} \ ,
		\label{gaugetrsf}
	\end{eqnarray}
	reflecting the fact that the charge world-lines and vortex world-surfaces they couple to are closed. Note that, at the classical level, the equations of motion imply that the gauge fields themselves encode the charge and vortex currents, respectively,
	\begin{eqnarray}
		Q_{\mu} = - {1\over 4\pi} \ell^3 k_{\mu \alpha \beta} b_{\alpha \beta} \ ,
		\nonumber \\
		M_{\mu \nu} = -{1\over 2\pi} \ell^2 k_{\mu \nu \alpha} a_{\alpha} \ .
		\label{eqmot}
	\end{eqnarray}
	If $a_{\mu}$ is a vector field and $b_{\mu \nu}$ a pseudotensor field, the model is also invariant under the parity ${\cal P}$ and time-reversal ${\cal T}$ symmetries. In general, the BF action for a model defined on a compact space with non-trivial topology has a ground state degeneracy\,\cite{semenoff} reflecting the topology, exactly as the Chern-Simons term in 2D. However, when the coefficient of the BF term is $1/4\pi$, as in (\ref{ac1}), the ground state is unique\cite{semenoff}. 
	
	Having established that the effective field theory for the inhomogeneous superconductors is a generalized gauge theory we can proceed as usual in its construction, adding order by order all interactions that respect the relevant symmetries. In this case, the next-order gauge-invariant terms are the kinetic terms for the two gauge fields. For the vector gauge field $a_{\mu}$ this is the usual Maxwell  term, constructed from the field strength $f_{\mu \nu} = d_{\mu} a_{\nu} -d_{\nu} a_{\mu}$. For the antisymmetric tensor gauge field $b_{\mu \nu}$, the kinetic term is quadratic in the field strength
	\begin{equation}
		h_{\mu \nu \alpha} = d_{\mu} b_{\nu \alpha}  + d_{\nu} b_{\alpha \mu} + d_{\alpha} b_{\mu \nu}\ .
		\label{kalb}
	\end{equation}
	It is easy to check that this is invariant under a gauge transformation (\ref{gaugetrsf}).
	Adding these next-order terms, we obtain the Euclidean effective action
	\begin{equation}
		S = \sum_x {\ell ^4\over 4f^2}
		f_{\mu \nu} f_{\mu \nu} +i{\ell ^4\over 4\pi }
		a_{\mu }k_{\mu \alpha \beta }b_{\alpha \beta } + {\ell ^4\over 12 \Lambda^2}
		h_{\mu \nu \alpha }h_{\mu \nu \alpha} 
		+i\ell a_{\mu }Q_{\mu }
		+i\ell ^2{1  \over 2} b_{\mu \nu}M_{\mu \nu} \ .
		\label{ac2}
	\end{equation}
	The dimensionless parameter $f={\cal O}(e)$ represents the effective Coulomb interaction strength in the material and $1/\Lambda$ is the magnetic screening length. The two dimensionless parameters $f$ and $\Lambda \ell$ encode the strengths of the electric and magnetic interactions, respectively. Non-relativistic effects can be easily incorporated by considering a velocity of light $v <1$ but are of no particular relevance for what follows. 
	
	We now integrate out the emergent gauge fields to obtain an effective (Euclidean) action for the charges and vortices alone,
	\begin{eqnarray}
		S_{\rm QM} = \sum_x {f^2 \over 2\ell^2} \ Q_{\mu }{\delta _{\mu \nu }
			\over {m^2- \nabla ^2 }} Q_{\nu }  \nonumber\\
		+ {\Lambda^2 \over 8} \ M_{\mu \nu}
		{{\delta _{\mu \alpha } \delta _{\nu \beta }-\delta _{\mu \beta}
				\delta _{\nu \alpha }} \over {m^2-\nabla ^2}} M_{\alpha \beta } 
		+ i {\pi m^2\over \ell} Q_{\mu} {k_{\mu \alpha \beta} \over \nabla^2 \left( m^2-\nabla^2\right)} M_{\alpha \beta} \ .
		\label{cvi}
	\end{eqnarray}
	where
	\begin{equation}
		m={f\Lambda\over 2\pi} \ ,
		\label{mass}
	\end{equation}
	is the gauge-invariant, topological mass, analogous to the famed Chern-Simons mass in 2D\,\cite{jackiw2, bowick}. This is one of the main points of this paper: the topological mutual statistics interaction screens both the vortex-vortex interaction and the Coulomb interaction between charges. Approximating these screened potentials by delta functions one can estimate the mass (coefficient multiplying the world-line length in the action) of charges and the string tension (coefficient multiplying the world-sheet area in the action) of vortices. 
	In this phase of the system both charges and vortices are gapped excitations with mass $M=f^2/(2m^2\ell^3)$ and string tension 
	$T=\Lambda^2/(8m^2\ell^2)$ interacting via short-range screened potentials. For temperatures low enough this is thus a thermally activated, insulating phase, for higher temperatures it is a metal. 
	
	Let us now investigate what happens when charges condense, which can be described by letting the original integer-valued variable $Q_{\mu}$ become a real-valued field over which one has to integrate (as opposed to sum) in the partition function. Formally, this amounts to using the Poisson summation formula
	\begin{equation}
		\sum_{\{Q_{\mu} \}} f\left( Q_{\mu} \right) = \sum_{\{k_{\mu} \}} \int dQ_{\mu} f\left( Q_{\mu} \right) {\rm e}^{i2\pi Q_{\mu} k_{\mu}} \ ,
		\label{poisson}
	\end{equation} 
	and focusing only on the sector in which the dual variable $k_{\mu } = 0$. However, since the current $Q_{\mu}$ is conserved and thus constrained by the equation $d_{\mu} Q_{\mu}=0$, we must first introduce the representation $Q_{\mu} =\ell k_{\mu \alpha \beta} n_{\alpha \beta}$. The new variables $n_{\alpha \beta}$ are now free but redundant, since they can be gauge transformed as in (\ref{gaugetrsf}). 
	There are only three gauge-invariant degrees of freedom in the $n_{\alpha \beta}$, corresponding to the three unconstrained variables $Q_{\mu}$. The removal of the three redundant variables can be taken care of by the usual gauge fixing in the integration. 
	
	We consider a 4D Euclidean lattice with spacing $\ell$ representing the typical distance of the superconducting droplets. Let $d_{\mu}$, $\hat d_{\mu}$, $S_{\mu}$ and $\hat S_{\mu}$ denote forward and backward lattice derivatives and shifts. Then the forward and backward lattice BF terms are defined by the three-index operators\cite{dst}
	\begin{eqnarray}
		k_{\mu \nu \rho} \equiv S_{\mu }\epsilon _{\mu \alpha
			\nu \rho } d _{\alpha } \ ,
		\nonumber \\
		\
		hat k_{\mu \nu \rho} \equiv \epsilon _{\mu \nu \alpha \rho}
		\hat d_{\alpha }\hat S_{\rho } \ ,
		\label{kop}
	\end{eqnarray}
	where no summation over equal indices $\mu$ and $\rho$ is implied. 
	The two lattice BF operators are interchanged (no minus sign) upon summation by parts on the lattice and are gauge invariant,
	\begin{eqnarray}
		k_{\mu \nu \rho} d _{\nu } = k_{\mu \nu \rho}
		d _{\rho } = \hat d_{\mu } k_{\mu \nu \rho} = 0 \ ,
		\nonumber \\
		\hat k_{\mu \nu \rho }d _{\rho } = \hat d _{\mu }
		\hat k_{\mu \nu \rho} = \hat d_{\nu }
		\hat k_{\mu \nu \rho } = 0 \ .
		\label{gaugeinv}
	\end{eqnarray}
	They also satisfy the identities 
	\begin{eqnarray}
		\hat k_{\mu \nu \rho} k_{\rho \lambda \omega} =
		-\left( \delta _{\mu \lambda} \delta_{\nu \omega} - \delta _{\mu \omega}
		\delta_{\nu \lambda } \right) \nabla^2  
		+ \left( \delta _{\mu \lambda }
		d_{\nu } \hat d_{\omega} - \delta _{\nu \lambda } d_{\mu }
		\hat d_{\omega } \right) \nonumber \\
		+ \left( \delta _{\nu \omega} d_{\mu }
		\hat d_{\lambda } - \delta _{\mu \omega} d_{\nu } \hat
		d_{\lambda } \right) \ ,
		\nonumber \\
		\hat k_{\mu \nu \rho} k_{\rho \nu \omega } = k_{\mu \nu \rho } \hat
		k_{\rho \nu \omega} = 2 \left( \delta _{\mu \omega } \nabla^2 - d_{\mu }
		\hat d_{\omega } \right) \ ,
		\label{maxwell}
	\end{eqnarray}
	where $\nabla^2 = \hat d_{\mu} d_{\mu}$ is the lattice Laplacian. 
	
	Using (\ref{maxwell}) we write the action (\ref{cvi}) 
	as 
	\begin{eqnarray}
		S_{\rm nM} = \sum_x \bigg[{\Lambda^2 \over 8} \ M_{\mu \nu}
		{{\delta _{\mu \alpha } \delta _{\nu \beta }-\delta _{\mu \beta} \delta _{\nu \alpha }} \over {m^2-\nabla ^2}} M_{\alpha \beta } 
		\nonumber \\
		+ {f^2 \over 2} n_{\mu \nu} {-\left( \delta _{\mu \alpha} \delta_{\nu \beta} - \delta _{\mu \beta}
			\delta_{\nu \alpha } \right) \nabla^2  
			+ \left( \delta _{\mu \alpha }
			d_{\nu } \hat d_{\beta} - \delta _{\nu \alpha } d_{\mu }
			\hat d_{\beta } \right) 
			\over m^2-\nabla^2} n_{\alpha \beta} \nonumber \\
		+ {f^2 \over 2} n_{\mu \nu}{ \delta _{\nu \beta} d_{\mu }
			\hat d_{\alpha } - \delta _{\mu \beta} d_{\nu } \hat
			d_{\alpha }  \over m^2-\nabla^2} n_{\alpha \beta} 
		\nonumber \\
		-i\pi  m^2 n_{\mu \nu} {\delta_{\mu \alpha} \delta_{\nu \beta} -\delta_{\mu \beta}
			\delta_{\nu \alpha} \over m^2-\nabla^2} M_{\alpha \beta}\bigg] \ .
		\label{scaction}
	\end{eqnarray}
	
	Performing Gaussian integration over the field $n_{\mu \nu}$, we obtain
	
	\begin{equation}
		S_{\rm M} = \sum_x {\Lambda^2 \over 8} \ M_{\mu \nu}
		{{\delta _{\mu \alpha } \delta _{\nu \beta }-\delta _{\mu \beta} \delta _{\nu \alpha }} \over -\nabla ^2} M_{\alpha \beta } \ .
		\label{long}
	\end{equation}
	This is the second main result of this paper. The global condensation in the superconducting phase turns the vortex interaction into a long-range one, suppressing the topological screening. As has been derived in \cite{dst}, a 4D Coulomb potential for the elements of the world-surface of a vortex implies that the self-energy of a circular vortex-loop of radius $r$ scales like $r\ {\rm ln} r$, which amounts to logarithmic vortex confinement as is the case in 2D. In these inhomogeneous 3D superconductors, thus, the destruction of global superconductivity via tunnelling percolation also takes places by vortex liberation, similarly to the 2D BKT transition. Since vortices are 1D objects, this happens when the effective string tension of the vortices, including the entropy correction, vanishes. This transition has been studied in \cite{dtv2} and leads to the VFT critical behaviour of the resistance given by  Eq.\,(\ref{vft}). 

	\section{Experiment}
	Standard four-terminal dc resistance measurements, see the inset in Fig.\,1a, are taken on films of the nitrides of the transition metals, NbTiN and NbN. The detailed preparation of the 
	20-nm-thick NbN film on an MgO substrate can be found in Methods. The zero-temperature coherence length of this film is measured to 
	be $(4.40 \pm 0.05)$ nm, i.e., much shorter than the film thickness $d=20$\,nm\, hence the NbN film is a 3D superconductor. 
	The details of the preparation of the disordered nonstoichiometric 86-nm-thick NbTiN film can be found in Ref.\,\cite{sheng}. This film is deposited on 
	a Si (100) substrate by radio frequency (RF) reactive magnetron co-sputtering from two separate NbN 
	and TiN targets. As the film is deposited on a Si substrate, it is fully compatible with the existing Si CMOS technology. 
	We benefit from the fact that this NbTiN film has been studied before\,\cite{sheng}. The zero-temperature coherence length is measured to 
	be $(9.53 \pm 0.04 )$ nm, which is much shorter than the film thickness 86 nm\,\cite{sheng}. Therefore, we also have a 3D superconductor system. Both samples 
	are approximately rectangular films with the length of $\approx$\,4.5\,mm and the width of $\approx$\,3.5\,mm. As shown later, we see that 
	the VFT model fits the data much better than the corresponding BKT one does. 
	
	Figure\,1a shows the four-terminal dc resistance measurements of the 20-nm-thick NbN film grown on the MgO substrate as a function of temperature $T$. 
	We observe a rather broad metal-superconductor transition with decreasing $T$. In order to further study this, we fit our experimental 
	results with the BKT- (red curve) and the VFT (blue curve) scaling resistivity.  Note that we use the same three fitting parameters, the critical temperature, the overall normalization of the resistance, and 
		a constant $b$ having the dimensionality of temperature) for the two fits. The aim is to compare the two scalings and identify the best one. Marking the six data points that significantly deviate from any 
	fit in grey, we see that the VFT dependence fits the experimental
	results much better than the BKT one does, see Fig.\,1a.  Analogous study is performed on the much thicker NbTiN film with a thickness of $86$\,nm deposited on a 
	Si substrate. Figure\,1b displays the same kind of the four-terminal dc resistance measurements. Again, we observe a broad metal-superconductor 
	transition with decreasing temperature. The BKT scaling clearly fails to fit the experimental data. The VFT fits describe the experimental data very well, except for 
	three points marked grey. While at present no concrete model for this deviation, which could be caused by nonstoichiometric disorder, can be offered, one can strongly assume that it is caused by 
		quantum corrections to the resistance which become essential in the close vicinity of the superconducting transition temperature $T_{\mathrm c}$ which noticeably exceeds $T_\mathrm{VFT}$. 
		The fact that in thicker (86 nm thick) NbTiN film these deviations are much smaller than in the 20 nm thick NbN film excellently agrees with this assumption. However, 
		detailed calculations beyond the scope of the present work are necessary to quantitatively explore this assumption.  
		
\begin{figure}
\includegraphics[width=1.0\linewidth]{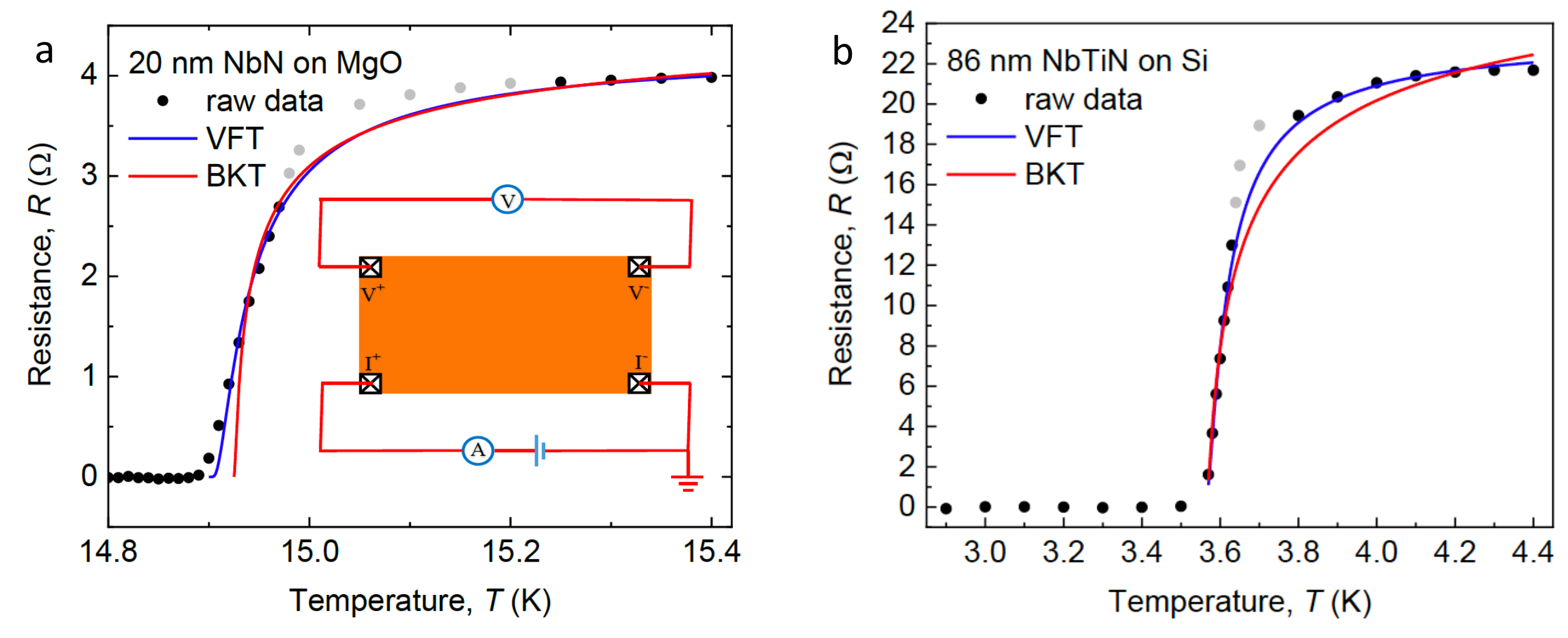}	
\caption{\textbf{Resistance measurements.}  \textbf{a:} Four-terminal dc resistance measurements of the 20-nm-thick NbN film at different temperatures.  The red and blue curves correspond to BKT fitting and VFT fitting to the experimental data, respectively. 
The gray points mark experimental data deviating from the fits. The inset depicts a sketch of the four-terminal dc resistance measurements of NbN and NbTiN films. Electrodes V$^+$ and V$^-$ correspond to the two voltage probes for measuring the voltage difference. Electrodes I$^+$ and I$^-$ are the source and drain contacts. The symbols V and I stand the voltmeter and ammeter measuring the voltage difference and current, respectively. \textbf{b:} Four-terminal dc resistance measurements of a 20-nm-thick NbN film at different temperatures. The red and blue curves correspond to the BKT fitting and VFT fitting to the experimental data, respectively. The three data points marked gray show the noticeable deviation from the fits.}\label{fig_pm_exp}
\end{figure}

	The next step is to confirm that the observed resistance behavior reflects the genuine bulk superconducting properties rather than stems 
	from the local superconductivity that might arise at the surface or 1D defect filaments of the investigated samples. To that end, we 
	perform magnetic susceptibility measurements. As shown in Fig.\,2, the investigated systems demonstrate strong diamagnetic response in the 
	magnetic susceptibility, evidencing that the observed superconductivity is the genuine bulk superconductivity for both films. The downward 
	then upward behavior of the magnetic susceptibility in the 86-nm-thick NbTiN film in the FC condition indicates possible paramagnetic 
	Meissner effect (PME) \,\cite{chen}. The PME generally appears in superconductors with strong vortex pinning at low temperatures.

\begin{figure}
		\includegraphics[width=1.0\linewidth]{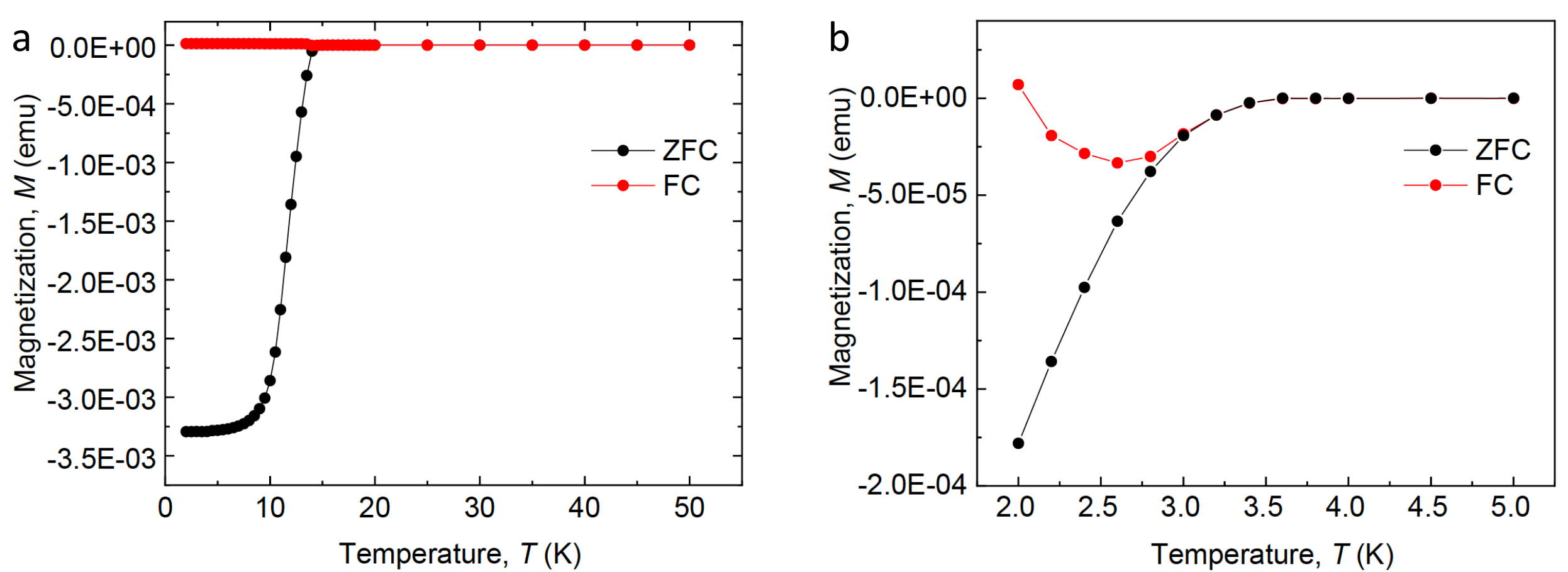}	
	\caption{\textbf{Magnetization measurements under zero-field-cooled (ZFC) and field-cooled (FC) condictions.} \textbf{a:}\,Magnetization of the 20\,nm-thick NbN film. Insert: magnetic susceptibility. The measurements are taken under the external field of 20 Oe.\textbf{b:}\,Magnetization of the 86\,nm-thick NbTiN film. The measurements are taken under the external field of 5 Oe.} \label{fig_pm_exp}
\end{figure}

	\section{Conclusion}
	Our results reveal the existence of a novel type of superconductivity, which we call type-III superconductivity. The standard superconductivity is well described by the local Ginzburg-Landau theory for a homogeneous, global order parameter in which gauge fields are screened by the Anderson-Higgs mechanism and which usually corresponds to the microscopic BCS pairing mechanism. The type III superconductivity is described by a topological gauge theory and corresponds to an inhomogeneous network of condensate droplets getting connected by tunneling pairs percolation and is destroyed by vortex liberation instead of pairs breaking. The underlying physics is the generalization of the BKT physics to three dimensions. The corresponding predicted modified exponential  VFT scaling of the resistance is fully confirmed by experiment. In 2D, only this second type of superconductivity survives due to strong infrared divergences in the Ginzburg-Landau theory, in 3D both types of superconductivity are possible. There are strong hints that the type of superconductivity that we describe here is associated with the BEC of strongly bound electron pairs and is realized in the high-$T_c$ materials. 
	
	
	\section{Experimental Section}
	\subsection{Preparation of the NbN film}
	We used the radio frequency (RF) reactive magnetron sputter to deposit a 20-nm-thick NbN film on the MgO (100) substrate in an 
	ultrahigh vacuum chamber with the base pressure of 3.9$\cdot 10^{-9}$\,Torr,
	We fixed the argon/nitrogen flow rate to 12:0.5. The fixed gas pressure of 3 mTorr and the fixed RF power of 120\,W are used. The argon plasma 
	(12 sccm) strikes the NbN target and atoms or molecules are ejected from the target surface. These atoms or molecules travel towards 
	the silicon substrate with the high temperature of 800 \,$^\circ$~C and deposit as the NbN film.

	\subsection{Preparation of the NbTiN film}
	The detailed description regarding the preparation of the disordered nonstoichiometric 86-nm-thick NbTiN film is given in\,\cite{sheng}. In short, our film is deposited on a Si (100) substrate by the RF reactive magnetron co-sputtering from two separate 
	NbN (99.5~\%) and TiN (99.5~\%) targets at 800 \,$^\circ$~C. 
	The base pressure in the chamber is less than 9$\cdot 10^{-9}$\,Torr, and the gas pressure is controlled at 3\,mTorr during the deposition. The RF 
	sputtering powers of both targets are set as 100\,W. The gas flow rate ratio of argon and
	nitrogen is 12:0.5. The argon plasma (12 sccm) strikes two targets, and atoms or molecules are ejected from the target's surface. 
	These atoms or molecules travel towards the silicon substrate with the high temperature of 800 \,$^\circ$~C and deposit as the NbTiN film. The large lattice mismatch between NbTiN and Si possibly leads to disordered and inhomogeneous nature of the NbTiN film.
	
	\subsection{Electrical measurements}
	The low-temperature four-terminal resistance measurements are performed in an Oxford Triton 200 cryo-free $^3$He/$^4$He dilution fridge. 
	We use the Keithley 2400 current source meter to provide a constant dc current that flows from the source to the drain contact. 
	On the other hand, A Keithley 2000 multimeter is used to measure the voltage drop between the two voltage probes. 
	
	\subsection{Magnetization measurements}
	Magnetic susceptibility measurements were carried out using a dc superconducting quantum interference device (SQUID) magnetometer. Both 
	zero-field-cooled (ZFC) and field-cooled (FC) regimes are used.

	\medskip
	\textbf{Acknowledgements} \par 
	The work of V.M.V. was supported by the Terra Quantum AG. Y.-J.L. and C.-T.L. would like to thank the National Science and Technology Council (NSTC), Taiwan for financial support.
	We thank Y.C. Yeh, H.C. Yeh, and J.W. Yang for their experimental help.
	
	\medskip
	
	%
	
	
	\bibliographystyle{plain}

\vskip 12pt	
	
	\begin{thebibliography}{10}
		\expandafter\ifx\csname url\endcsname\relax
		\def\url#1{\texttt{#1}}\fi
		\expandafter\ifx\csname urlprefix\endcsname\relax\def\urlprefix{URL }\fi
		\providecommand{\bibinfo}[2]{#2}
		\providecommand{\eprint}[2][]{\url{#2}}
		
		
	
		\bibitem{tinkham} 
		M.\,Tinkham, Introduction to superconductivity, Dover Publications, New York 1996. 
		
		\bibitem{granular} 
		D.\,Kowal, Z.\,Ovadyahu,
			Solid St. Comm. 1994, 90, 783.
			
	\bibitem{bat_vin_2007}
	T.\,I.\,Baturina, A.\,Yu.\,Mironov, V.\,M.\,Vinokur, M.\,R.\,Baklanov, C.\,Strunk,
	Phys. Rev. Lett. 2007, 99, 257003.
				
				\bibitem{dst}
				M.\,C.\,Diamantini, P.\,Sodano, C.\,A.\,Trugenberger,
				Nuclear Physics, 1996 B474, 641.
				
				\bibitem{vinokurnature}
				V.\,M.\,Vinokur, T.\,I.\,Baturina, M.\,V.\,Fistul, A.\,Yu.\,Mironov, M.\,R.\,Baklanov, C.\,Strunk, 
				Nature, 2008, 452, 613. 
				
				\bibitem{dtv1}
				M.\,C.\,Diamantini,\,M.\,C., Trugenberger,\,C.\,A. \& Vinokur,\,V.\,M. 
					Comm. Phys. 2018, 1, 77. 
					
					\bibitem{das1} 
					D.\,Das, S.\,Doniach,
						Phys. Rev. 1999, B 60, 1261.
						
						\bibitem{bm}
						M.\,C.\,Diamantini, A.\,Yu.\,Mironov, S.\,V.\,Postolova, X.\,Liu, Z.\,Hao, D.\,M.\,Silevitch, Ya.\,Kopelevich, P.\,Kim, C.\,A.\,Trugenberger, V.\,M.\,Vinokur, 
							Phys. Lett. 2020, 384, 126570.
							
							\bibitem{dtv-SIT}
							M.\,C.\,Diamantini, L.\,Gammaitoni, C.\,A.\,Trugenberger, V.\,M.\,Vinokur, 
								J. Supercond. Nov. Magn. 2019, 32, 47. 
								
								\bibitem{book1} 
								M.\,C.\,Diamantini, C.\,A.\,Trugenberger, V.\,M.\,Vinokur,
									in Topological Phase Transitions and New Developments,  World Scientific, Singapore, 197, 2022. 
									
									\bibitem{book}
									C.\,A.\,Trugenberger, Superinsulators, Bose metals and high-$T_c$ superconductors: the quantum physics of emergent magnetic monopoles. World Scientific, Singapore, 2022. 
									
									\bibitem{sacepe1} 
									B.\,Sac\'ep\'e, T.\,I.\,Baturina, V.\,M.\,Vinokur, M.\,R.\,Baklanov, M.\,Sanquer. 
										Phys. Rev. Lett. 2008, 101 157006. 
								
\bibitem{sacepe2}
B.\,Sac\'ep\'e, T.\,Dubouchet, C.\,Chapelier, M.\,Sanquer, M.\,Ovadia, D.\,Shahar, M.\,Feigel’man, L.\,Ioffe, 
Nature Physics 2011, 7, 239. 
											
											
\bibitem{polyakov}
A.\,M.\,Polyakov, 
Phys. Lett. 1975, 59, 82.  
												
\bibitem{templeton}
R.\,Jackiw, S.\,Templeton, 
Phys. Rev.\,D\,1981, 23, 2291. 
													
\bibitem{vinokur2011}
A.\,Glatz, I.\,S.\,Aranson, T.\,I.\,Baturina, N.\,M.\,Chtchelkatchev, V.\,M.\,Vinokur, 
Phys. Rev.\,B\,2011, 84, 024508.
													
\bibitem{infrared}
M.\,C.\,Diamantini, C.\,A.\,Trugenberger, V.\,M.\,Vinokur, 
Journal of High Energy Physics, 2022, 10, 100.
													
\bibitem{3Dgranular} 
C.\,Parra, F.\,Niemstemski, A.\,W.\,Contryman, P.\,Giraldo-Gallo, T.\,H\,Geballe, I.\,R.\,Fisher, H.\,C.\,Manoharan, 
PNAS, 2021, 118, e2017810118. 
														
\bibitem{muller} 
D.\,Mihailovic, V.\,V.\,Kabanov, K.\,A.\,M\"uller, 
Europhys. Lett. 2002, 57, 254. 
															
\bibitem{barisic} 
D.\,Pelc, M.\,Vuckovic, M.\,S.\,Grbic, Yu.\,G.\,Pozek, T.\,Sasagawa, M.\,Greven, N.\,Barisi\'{c}, 
Nature Comm. 2018, 9 4327. 
																
\bibitem{pairs2D}
K.\,M.\,Bastiaans, D.\,Chatzopoulos, J.-F.\,Ge, D.\,Cho, W.\,O.\,Tromp, J.\,M.\,van\,Ruitenbeek, M.\,H.\,Fischer, P.\,J.\,de\,Visser, D.\,J.\,Thoen, E.\,F.\,C.\,Driessen, T.\,M.\,Klapwijk, M.\,P.\,Allan, 
Science,\,2021, 374, 608.
																
\bibitem{pairs3D} 
P.\,Zhou, L.\,Chen, Y.\,Liu, I.\,Sochnikov, A.\,T.\,Bollinger, M.-G.\,Han, Y.\,Thu, I.\,Bozovic, D.\,Natelson, 
Nature, 2019, 572, 493. 
																	
\bibitem{SIT3D} 
A.\,T.\,Bollinger, G.\,Dubuis, J.\,Yoon, D.\,Pavuna, J.\,Misewich, I.\,Bozovic, 
Nature, 2011, 472, 458. 
																		
\bibitem{bec} 
Z.\,H.\,Luo, W.\,Pang, B.\,Liu, Y.-Y.\,Li, B.\,A.\,Malomed, 
Frontiers of Physics, 2021, 16, 32201. 
																						
\bibitem{wilczek}
F.\,Wilczek, 
Phys. Rev. Lett. 1992, 69, 132. 

\bibitem{para} A.\, P.\, Nielsen et al. {\it Phys. Rev}. B\,2000, 62, 14380.
																						
\bibitem{topsc}
M.\,C.\,Diamantini, P.\,Sodano, C.\,A\,Trugenberger, 
Eur. Phys. J. 2016,  B 53, 19.
																					
\bibitem{higgsless} 
M.\,C.\,Diamantini, C.\,A\,Trugenberger, 
Nucl. Phys. 2015, 891, 401.
																						
\bibitem{minnhagen} 
P.\,Minnhagen, 
Rev. Mod. Phys. 1987, 59, 1001. 	
																							
\bibitem{dtv2}
	M.\,C.\,Diamantini, L.\,Gammaitoni, C.\,A\,Trugenberger, V.\,M.\,Vinokur, 
Scientific Reports, 2018, 8, 15718. 
																					
\bibitem{shahar}
M.\,Ovadia, D.\,Kalok, I.\,Tamir, S.\,Mitra, B.\,Sacep\'e, D.\,Shahar, 
Scientific Reports, 2015, 5, 13503. 
																									
\bibitem{vasin}
M.\,G.\,Vasin, V.\,N.\,Ryzhov, V.\,M.\,Vinokur, 
arXiv:1712.00757, 2017.
																									
\bibitem{disBKT}
S.\,Sankar, V.\,M.\,Vinokur, V.\,Tripathi, 
Phys. Rev. B\,2018, 97, 020507(R).
																										
\bibitem{moncon}
M.\,C.\,Diamantini, C.\,A\,Trugenberger, V.\,M.\,Vinokur, 
Communications Physics, 2021, 4, 25.
																											
\bibitem{bohm} 
Y.\,Aharonov, D.\,Bohm, 
Phys. Rev. 1961, 115, 485.
																												
\bibitem{casher}
Y.\,Aharonov, A.\,Casher, 
Phys. Rev. Lett. 1984, 53, 319.
																													
\bibitem{kaufmann}
L.\,H.\,Kaufmann, Formal\,knot\,theory, Princeton University Press, Princeton, 1983. 
																													
\bibitem{jackiw2} 
S.\,Deser, R.\,Jackiw, S.\,Templeton, 
Phys. Rev. Lett. 1982, 48, 975.
																													
\bibitem{bowick} 
T.\,J.\,Allen, M.\,J.\,Bowick, A.\,Lahiri, 
Mod. Phys. Lett. A, 1991, 6, 559.
																														
\bibitem{semenoff}
M.\,Bergeron, G.\,W.\,Semenoff, R.\,J.\,Szabo, 
Nucl. Phys.\,B, 1995,\,437, 695.
																															
\bibitem{sheng} 
S.-Z.\,Chen, J.-W.\,Yang, T.-Y.\,Peng, Y.-C.\,Chu, C.-C.\,Yeh, I.-F.\,Hu, S.\,Mhatre, Y.-J.\,Lu, C.-T.\,Liang, 
Supercond. Sci. Technol. 2022, 35, 064003. 
																																
\bibitem{chen} 
F.\,H.\,Chen, M.\,F.\,Tai, W.\,C.\,Horng, T.\,Y.\,Tseng, 
Phys. Rev. B, 1993, 48, 1258. 

\end{thebibliography}

\end{document}